\documentclass[preprint, 5p]{elsarticle}

\usepackage{amsmath,amssymb,bm}
\usepackage{lineno}
\usepackage{color}

\usepackage{multirow}
\usepackage{multicol}

\journal{Physics Letters B}

\begin{document}

\begin{frontmatter}

\title{Relativistic model-free prediction for neutrinoless double beta decay at leading order}

\author[PKU]{Yilong Yang}
\author[PKU]{Pengwei Zhao\corref{mycorrespondingauthor}}
\ead{pwzhao@pku.edu.cn}
\address[PKU]{State Key Laboratory of Nuclear Physics and Technology, School of Physics, Peking University, Beijing 100871, China}

\cortext[mycorrespondingauthor]{Corresponding authors}

\begin{abstract}
Starting from a manifestly Lorentz-invariant chiral Lagrangian, we present a model-free prediction for the transition amplitude of the process $nn\rightarrow pp e^-e^-$ induced by light Majorana neutrinos, which is a key process of the neutrinoless double beta decay ($0\nu\beta\beta$) in heavy nuclei employed in large-scale searches.
Contrary to the nonrelativistic case, we show that the transition amplitude can be renormalized at leading order without any uncertain contact operators.
The predicted amplitude defines a stringent benchmark for the previous estimation with model-dependent inputs, and greatly reduces the uncertainty of  $0\nu\beta\beta$ transition operator in the calculations of nuclear matrix elements.
Generalizations of the present framework could also help to address the uncertainties in $0\nu\beta\beta$ decay induced by other mechanisms.
In addition, the present work motivates a relativistic {\it ab initio} calculation of $0\nu\beta\beta$ decay in light and medium-mass nuclei.
\end{abstract}

\begin{keyword}
Neutrinoless double-beta decay \sep Nuclear matrix elements \sep Decay operator \sep Relativistic effects
\end{keyword}

\end{frontmatter}

\section{Introduction}
Neutrinoless double beta decay ($0\nu\beta\beta$) is a second-order weak process, in which a nucleus decays to its neighboring nucleus by turning two neutrons to two protons, emitting two electrons but no corresponding antineutrinos~\cite{Furry1939Phys.Rev.11841193}.
Its observation would provide direct evidence of lepton number violation beyond the standard model, prove the Majorana nature of neutrinos~\cite{Schechter1982Phys.Rev.D29512954,Zeldovich1981JETPLett.141}, constrain the neutrino mass scale and hierarchy~\cite{Avignone2008Rev.Mod.Phys.481}, and shed light on the matter-antimatter asymmetry in the Universe~\cite{Davidson2008Phys.Rep.105}.
Therefore, it becomes one of the top priorities in the field of nuclear and particle physics, and invigorates worldwide experiments either activate or planned~\cite{Aalseth2018Phys.Rev.Lett.132502,Adams2020Phys.Rev.Lett.122501,Agostini2020Phys.Rev.Lett.252502,
Albert2018Phys.Rev.Lett.072701,Armengaud2021Phys.Rev.Lett.181802,Arnold2017Phys.Rev.Lett.041801,Gando2016Phys.Rev.Lett.082503,Dai2022Phys.Rev.D32012} (for a recent review, see Ref.~\cite{Agostini2023Rev.Mod.Phys.025002}).

Nuclear matrix element, which encodes the impact of the nuclear structure on the decay half-life, is crucial to interpreting the experimental limits and even more potential future discoveries.
Within the standard picture of $0\nu\beta\beta$, which involves the long-range light neutrino exchange~\cite{Weinberg1979Phys.Rev.Lett.15661570}, a minimal extension of the standard model, current knowledge of the nuclear matrix element is not satisfactory~\cite{Engel2017Rep.Prog.Phys.046301}, as various nuclear models lead to discrepancies as large as a factor around 3.

Chiral effective field theory (EFT)~\cite{Weinberg1990Phys.Lett.B288,Weinberg1991Nucl.Phys.B318} plays an important role in addressing such uncertainties.
It can provide the nuclear Hamiltonian and weak currents in a consistent and systematically improvable manner~\cite{Epelbaum2009Rev.Mod.Phys.17731825, Machleidt2011Phys.Rep.175, Hammer2020Rev.Mod.Phys.025004}.
Based on chiral EFT, the issue of $g_A$ quenching in single $\beta$ decays~\cite{Towner1987PhysicsReports263377} was resolved as a
combination of two-nucleon weak currents and strong nuclear correlations~\cite{Pastore2018Phys.Rev.C022501,Gysbers2019NaturePhysics428431}.
The chiral-EFT-based $0\nu\beta\beta$ transition operators under various sources of lepton number violation~\cite{Prezeau2003Phys.Rev.D034016,Cirigliano2017J.HighEnergyPhys.82,Cirigliano2018Phys.Rev.C065501,Cirigliano2018Phys.Rev.Lett.202001,
Cirigliano2019Phys.Rev.C055504,Cirigliano2021Phys.Rev.Lett.172002,Cirigliano2021J.HighEnergyPhys.289},
as well as their impacts on the nuclear matrix elements have been extensively studied in the literature~\cite{Menendez2011Phys.Rev.Lett.062501,Pastore2018Phys.Rev.C014606,Wang2018Phys.Rev.C031301,Jokiniemi2021Phys.Lett.B136720,Yao2021Phys.Rev.C014315, Weiss2022Phys.Rev.C065501}.
In addition, {\it ab initio} calculations of nuclear matrix elements using a chiral Hamiltonian are available recently~\cite{Yao2020Phys.Rev.Lett.232501, Belley2021Phys.Rev.Lett.042502, Novario2021Phys.Rev.Lett.182502, Wirth2021Phys.Rev.Lett.242502}.

In the context of the standard light Majorana neutrino-exchange mechanism~\cite{Weinberg1979Phys.Rev.Lett.15661570}, Cirigliano {\it et al.} showed that a chiral EFT description of $0\nu\beta\beta$ decay, already at leading order (LO), requires a contact operator to ensure renormalizability~\cite{Cirigliano2018Phys.Rev.Lett.202001, Cirigliano2019Phys.Rev.C055504}.
The size of this contact term should be determined in principle by matching to first-principle gauge field theory calculations, which however are not yet available currently~(see Refs.~\cite{Feng2019Phys.Rev.Lett.022001,Tuo2019Phys.Rev.D094511,Cirigliano2020Prog.Part.Nucl.Phys.103771,
Detmold2020arXiv2004.07404,Feng2021Phys.Rev.D034508,Davoudi2021Phys.Rep.174,Davoudi2021Phys.Rev.Lett.152003,Tuo2022Phys.Rev.D074510,Davoudi2022Phys.Rev.D094502} for related progress).
This unknown contact term, thus, leads to an additional source of uncertainty for the nuclear matrix elements besides the nuclear-structure ones.

Cirigliano {\it et al.}~\cite{Cirigliano2021Phys.Rev.Lett.172002,Cirigliano2021J.HighEnergyPhys.289} first modeled the $nn\rightarrow pp e^-e^-$ transition amplitude with an integral representation, i.e., a momentum integral of the neutrino propagator times the generalized forward Compton scattering amplitude.
This approach introduces, in the intermediate-momentum region, model-dependent inputs from elastic intermediate states in analogy to the Cottingham formula~\cite{Cottingham1963Ann.Phys.N.Y.424432}.
In the absence of experimental datum and costly calculations based on lattice gauge theory, the obtained model-dependent amplitude could be used as a pseudo datum to determine the contact term in nuclear structure calculations~\cite{Wirth2021Phys.Rev.Lett.242502}. However, a model-free prediction of the transition amplitude is still missing.

In this Letter, we present a model-free prediction of the $nn\rightarrow pp e^-e^-$ amplitude by developing a relativistic framework based on chiral EFT.
The term ``model-free" here means the obtained results are free of model-dependent inputs beyond the framework of the chiral EFT.
We show that the transition amplitude can be renormalized at LO without the vague contact term in the relativistic framework starting from a manifestly Lorentz-invariant chiral Lagrangian.
Matching to our results will allow one to stringently determine the contact term needed in nonrelativistic nuclear-structure calculations, and to greatly reduce the corresponding uncertainty in the calculations of nuclear matrix elements for $0\nu\beta\beta$ decay.
It also paves the way to relativistic {\it ab initio} calculations of $0\nu\beta\beta$ decay rates in light and medium-mass nuclei.

\section{Relativistic framework}
We start from a manifestly Lorentz-invariant effective Lagrangian relevant at the leading order of chiral EFT~\cite{Machleidt2011Phys.Rep.175},
\begin{equation}
  \begin{split}
    \mathcal{L}_{\Delta L=0}&=\frac{f_\pi^2}{4}{\rm tr}[u_\mu u^\mu+m_\pi^2(uu+u^\dagger u^\dagger)]\\
    &+\overline{\Psi}({\rm i}\gamma^\mu D_{\mu}-M+\frac{g_A}{2}\gamma^\mu\gamma_5 u_\mu)\Psi-\sum_\alpha\frac{C_\alpha}{2}(\overline{\Psi}\Gamma_\alpha\Psi)^2,
  \end{split}
\end{equation}
where $f_\pi=92.2$ MeV is the pion decay constant, $g_A=1.27$ is the nucleon axial coupling, $M$ denotes the nucleon mass, and $C_\alpha$ $(\alpha=S,V,A,AV,T)$ are the low-energy constants (LECs).
This Lagrangian consists of the pion field $u=\exp[{\rm i}\vec{\tau}\cdot\vec{\pi}/(2f_\pi)]$ and the nucleon field $\Psi=(p,n)^T$, which are coupled to the weak current $l_\mu$ via the axial vector $u_\mu={\rm i}u^\dagger(\partial_\mu -{\rm i}l_\mu)u-{\rm i}u\partial_\mu u^\dagger$ and the chirally covariant derivative $D_\mu=\partial_\mu+\frac{1}{2}[u^\dagger(\partial_\mu-{\rm i}l_\mu)u+u\partial_\mu u^\dagger]$.
The weak current reads $l_\mu=-2\sqrt{2}G_F V_{ud}\tau^+ \overline{e}_L\gamma_\mu\nu_{eL}+{\rm h.c.}$, with the Fermi constant $G_F$ and the $V_{ud}$ element of the Cabibbo-Kobayashi-Maskawa (CKM) matrix~\cite{Cabibbo1963Phys.Rev.Lett.531533,Kobayashi1973Prog.Theor.Phys.652657}.

For the standard mechanism of $0\nu\beta\beta$ decay, the lepton number violation at low energy is dominated by the electron-neutrino Majorana mass
\begin{equation}
  \mathcal{L}_{\Delta L=2}=-\frac{m_{\beta\beta}}{2}\nu^T_{eL} C \nu_{eL},
\end{equation}
where $C={\rm i}\gamma_2\gamma_0$ denotes the charge conjugation matrix, and $m_{\beta\beta}$ the effective neutrino mass.

Contrary to the previous studies based on the heavy baryon approach~\cite{Cirigliano2018Phys.Rev.Lett.202001, Cirigliano2019Phys.Rev.C055504}, which relies on a nonrelativistic expansion of the Lagrangian, here we apply the manifestly Lorentz-invariant Lagrangian to the problem of $0\nu\beta\beta$ decay.
The LO contribution to the scattering amplitude can be obtained by solving the relativistic scattering equation
\begin{equation}\label{Eq.BS}
  \begin{split}
    T(\bm p',\bm p)&=V(\bm p',\bm p) \\
    &+\int\frac{{\rm d}^3 k}{(2\pi)^3}
   \frac{M^2}{\bm k^2+M^2}
  \frac{V(\bm p',\bm k)T(\bm k,\bm p)}{E-2\sqrt{\bm k^2+M^2}+{\rm i}0^+},
  \end{split}
\end{equation}
where $E$ is the total energy, and $\bm p'$ and $\bm p$ are the nucleon outgoing and incoming momenta in the center of mass frame, respectively.

Equation (\ref{Eq.BS}) is consistent with the three-dimensional reduction of the Bethe-Salpeter equation~\cite{Woloshyn1973Nucl.Phys.B269} and it satisfies relativistic elastic unitarity.
The LO potential $V=\overline{\varphi}_0\otimes\overline{\varphi}_0\mathcal{V}\varphi_0\otimes \varphi_0$ is defined by the LO two-nucleon irreducible diagrams   sandwiched between the leading term of the Dirac spinor $\varphi(\bm p,s)$ expanding in powers of small momenta $\bm p$.
The subleading terms of the Dirac spinor and the retardation effects on pion and neutrino exchanges are to be included perturbatively at higher orders.
As a result, the LO strong and neutrino potentials take the form as those in the Weinberg's approach,
\begin{eqnarray}
    \label{eq.Vs}
    V_{s}(\bm p',\bm p)&=&-\frac{g_A^2}{4f_\pi^2}\vec{\tau}_1\cdot\vec{\tau}_2\frac{\bm\sigma_1\cdot\bm q\bm\sigma_2\cdot\bm q}{\bm q^2+m_\pi^2}+C_1+C_2\bm\sigma_1\cdot\bm\sigma_2,\\
    \label{eq.Vnu}
    V_\nu(\bm p',\bm p)&=&\frac{\tau_1^+\tau_2^+}{\bm q^2}\left[1-g_A^2
    \bm\sigma_1\cdot\bm\sigma_2\right.\nonumber\\
    &+&\left.g_A^2\bm\sigma_1\cdot\bm q\bm\sigma_2\cdot\bm q\frac{2m_\pi^2+\bm q^2}{(\bm q^2+m_\pi^2)^2}\right],
\end{eqnarray}
where $\bm q=\bm p'-\bm p$, and $C_1=C_S+C_V$ and $C_2=-C_{AV}+2C_T$ are two independent LECs.

Note that the present derivation is similar to the so-called modified Weinberg approach~\cite{Epelbaum2012Phys.Lett.B338344}, which was applied to nucleon-nucleon scattering problem.
It has proven to be useful to improve the renormalizability of nucleon-nucleon scattering~\cite{Epelbaum2012Phys.Lett.B338344} and few-body systems~\cite{Epelbaum2017Eur.Phys.J.A98,Yang2022Phys.Lett.B137587}.
In the heavy baryon approach, the nonrelativistic expansion of the Lagrangian leads, instead of Eq.~(\ref{Eq.BS}), to the Lippmann-Schwinger equation, which contains a nonrelativistic limit $(1/M\rightarrow0)$ of the two-nucleon propagator
\begin{equation}\label{eq.prop}
  \frac{M^2}{\bm k^2+M^2}
  \frac{1}{E-2\sqrt{\bm k^2+M^2}+{\rm i}0^+}\rightarrow\frac{1}{E_{\rm kin}-\bm k^2/M+{\rm i}0^+}.
\end{equation}
Obviously, the relativistic propagator scales as $\mathcal{O}(\Lambda^{-3})$ at the ultraviolet (UV) region $|\bm k|\sim\Lambda$, while the nonrelativistic one scales as $\mathcal{O}(\Lambda^{-2})$.
As a result of this milder UV behavior, we will show that the $0\nu\beta\beta$ amplitude can be renormalized without promoting a contact term to the LO neutrino potential.

\section{Renormalization of the $0\nu\beta\beta$ amplitude}
\begin{figure}[!htpb]
    \centering
	\includegraphics[width=0.45\textwidth]{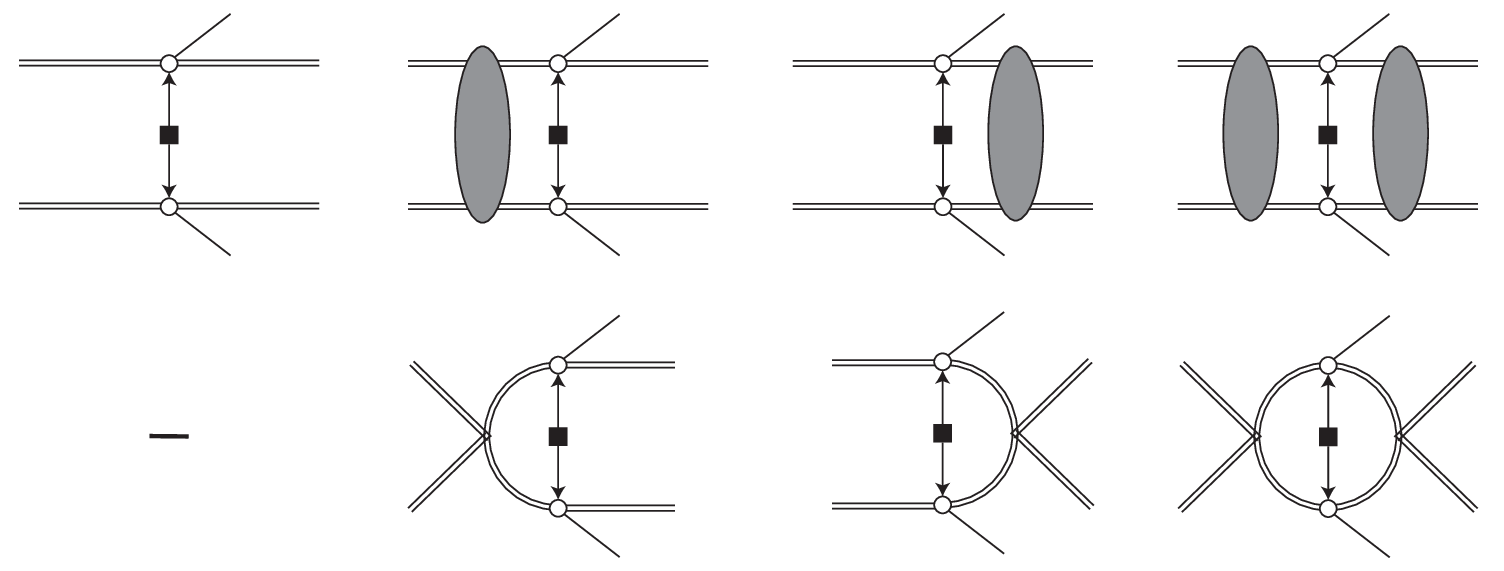}
	\caption{
    Leading-order contributions to the amplitude of $nn\rightarrow ppe^-e^-$ (first row) and the corresponding ultraviolet divergence structures (second row).
    The double and plain lines denote nucleon and lepton fields, respectively.
    The squares denote an insertion of neutrino potential $V_\nu$.
    The circles denote the nucleon axial and vector currents coupled to $V_\nu$.
    The gray ellipses represent the $T$ matrix generated by iteration of the strong potential $V_s$.
	}\label{fig1}
\end{figure}

We focus on the scattering process $nn\rightarrow ppe^-e^-$ in the $^1S_0$ wave, as it is the only channel that requires a contact term to ensure the renormalizability in the heavy-baryon approach~\cite{Cirigliano2018Phys.Rev.Lett.202001,Cirigliano2019Phys.Rev.C055504}.
Apart from the $^1S_0$ channel, there are other channels with nonzero angular momenta that have fairly large contributions to the nuclear matrix elements in nuclear structure calculations~\cite{Siimkovic2008Phys.Rev.C045503,Fang2018Phys.Rev.C045503},
but the LO amplitude can be renormalized in these channels without the contact terms.

Without loss of generality for our arguments, we consider the kinematics $n(\bm p_i)n(-\bm p_i)\rightarrow p(\bm p_f)p(-\bm p_f)e(\bm p_{e1}=0)e(\bm p_{e2}=0)$.
The LO amplitude can be schematically written as
\begin{equation}\label{eq.amplitude}
  \mathcal{A}^{\rm LO}_\nu=-\rho_{fi}(V_\nu+V_\nu G_0 T_s+T_s G_0 V_\nu+T_s G_0 V_\nu G_0 T_s),
\end{equation}
where $\rho_{fi}=M^2/\sqrt{(\bm p_f^2+M^2)(\bm p_i^2+M^2)}$ is a phase space factor~\footnote{The phase factor $\rho_{fi}$ is determined by the unitarity.
The amplitude $\mathcal{A}_\nu$ is related to the $S$-matrix element by
$S_\nu={\rm i}(2\pi)^4\delta^{(4)}(P_f-P_i)4G_F^2 V_{ud}^2 m_{\beta\beta}\overline{u}_{eL}C\overline{u}^T_{eL} A_\nu$, consistent with Ref.~\cite{Cirigliano2018Phys.Rev.Lett.202001}.}, $G_0$ the two-nucleon free propagator, and $T_s$ the $T$-matrix resumming the strong potential $V_s$.
The four terms in Eq.~(\ref{eq.amplitude}) correspond to the four diagrams depicted in the first row of Fig.~\ref{fig1}, and here we denote them as
$\mathcal{A}_A$, $\mathcal{A}_B$, $\overline{\mathcal{A}}_B$, and $\mathcal{A}_C$ from left to right.

To study the renormalization of the transition amplitude, we now discuss the degree of UV divergence for the diagrams $\mathcal{A}\sim\mathcal{O}(\Lambda^D)$ ($\log\Lambda$ for $D=0$).
The $\mathcal{A}_A$ tree diagram has no divergence.
For $\mathcal{A}_B$, $\overline{\mathcal{A}}_B$, and $\mathcal{A}_C$, it was shown that the degree of divergence is dominated by the loop integrals involving the insertion of neutrino potential, once $T_s$ is made finite~\cite{Cirigliano2018Phys.Rev.Lett.202001}.
The divergence structures are shown in the second row of Fig.~\ref{fig1}.
Counting the powers of loop momenta, one finds the degree of divergence
\begin{equation}\label{eq.divergence}
  D=L(3+g)-2
\end{equation}
with $L$ being the number of loops and $g$ the UV scaling of the two-nucleon propagator, and $-2$ comes from the $|\boldsymbol q|^{-2}$ dependence of $V_\nu$.

From Eq.~(\ref{eq.prop}), we know the nonrelativistic propagator has $g=-2$ and the relativistic one has $g=-3$.
Therefore, in the nonrelativistic framework, $\mathcal{A}_B$ and $\overline{\mathcal{A}}_B$ are convergent as $\mathcal{O}(\Lambda^{-1})$, but $\mathcal{A}_C$ is logarithmic divergent.
In the present relativistic framework, however, $\mathcal{A}_B$, $\overline{\mathcal{A}}_B$, and $\mathcal{A}_C$ are all convergent as $\mathcal{O}(\Lambda^{-2})$, so  the corresponding LO amplitude is renormalizable.
Such analysis should hold true not only for chiral EFT, but also for pionless EFT, since the dominating UV divergence structure does not involve pion exchanges.

The renormalizability of the LO amplitude $\mathcal{A}^{\rm LO}_\nu$ can be demonstrated explicitly in any specific regularization scheme.
Here, we regulate the strong potential with a separable gaussian function,
\begin{equation}
  V_s(\bm p',\bm p)\rightarrow{\rm e}^{-|\bm p'|^4/\Lambda^4}V_s(\bm p',\bm p){\rm e}^{-|\bm p|^4/\Lambda^4}.
\end{equation}
After projecting to the $^1S_0$ channel, the LEC in its short-range part $C^{^1S_0}=C_1-3C_2$ is determined by reproducing the scattering length $a_{np}=-23.74$ fm.
We have checked that the $^1S_0$ phase shifts are indeed cutoff independent as $\Lambda\rightarrow\infty$ .
Then, the amplitude is evaluated with Eq.~(\ref{eq.amplitude}) in the momentum space.

\begin{figure}[!htpb]
    \centering
	\includegraphics[width=0.45\textwidth]{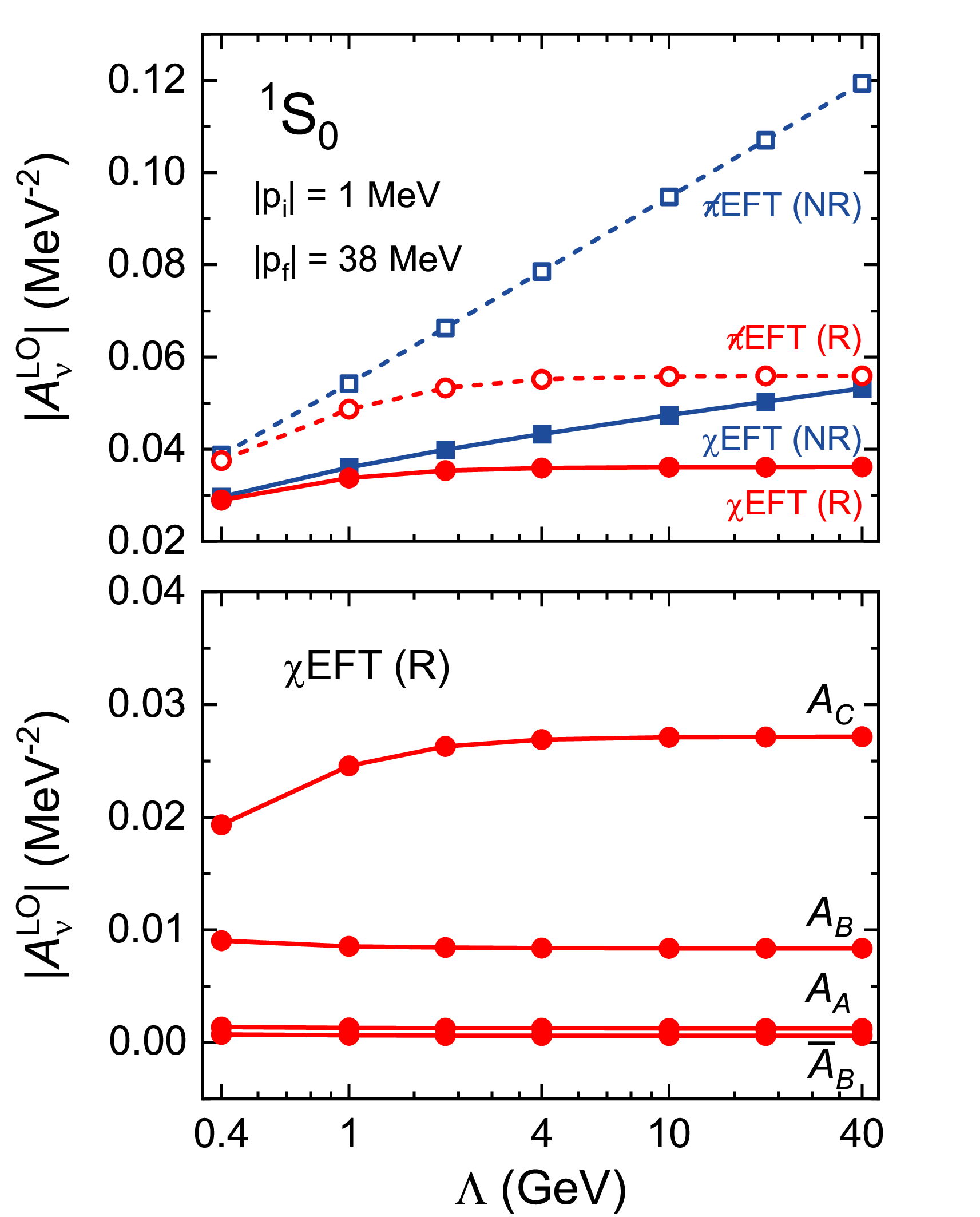}
	\caption{(Color online). The upper panel depicts the amplitudes $\mathcal{A}^{\rm LO}_\nu$ for $|\bm p_i|=1$ MeV and $|\bm p_f|=38$ MeV, as functions of the cutoff $\Lambda$.
The solid and open circles (squares) denote the results of the relativistic (nonrelativistic) chiral and pionless EFTs, respectively.
The lower panel depicts the leading-order contributions to the amplitude (see the first row of Fig.~\ref{fig1}) obtained with the relativistic chiral EFT.
}\label{fig2}
\end{figure}

Taking the kinematics $|\bm p_i|=1$ MeV and $|\bm p_f|=38$ MeV as an example, the LO amplitude $\mathcal{A}^{\rm LO}_\nu$ is plotted against the cutoff $\Lambda$ in the upper panel of Fig.~\ref{fig2}.
A logarithmic divergence of the amplitudes can be clearly seen for the results given by the nonrelativistic pionless and chiral EFTs, consisting with the analysis above.
This leads to the introduction of a contact term to ensure the renormalizability, as proposed in Ref.~\cite{Cirigliano2018Phys.Rev.Lett.202001}.
In contrast, the amplitudes obtained in the relativistic framework, both chiral and pionless EFTs, converge against the increasing cutoff.
This means that the amplitudes can be predicted at LO without any undetermined LECs at all kinematics.

In the lower panel of Fig.~\ref{fig2}, the leading-order contributions to the amplitude (see the first row of Fig.~\ref{fig1}) are depicted separately, as functions of the cutoff.
The two-loop diagram $\mathcal{A}_C$ dominates the cutoff dependence of the full LO amplitude in the present relativistic framework, while it is logarithmic divergent in the nonrelativistic case.

Note that recent studies on the renormalization of chiral EFT in a scheme with finite cutoffs have been applied to nucleon-nucleon scattering~\cite{Gasparyan2022Phys.Rev.C024001,Gasparyan2023Phys.Rev.C044002}, and extending such a scheme to $0\nu\beta\beta$ decay process could also be interesting in the future, and the present results will provide a benchmark for such studies.

One can define the LO $0\nu\beta\beta$ operator for calculating nuclear matrix elements.
In the momentum space, it is defined as $V^{\rm LO}_\nu=-\mathcal{A}_A$ with $\mathcal{A}_A$ the tree-level amplitude~\cite{Cirigliano2018J.HighEnergyPhys.97},
\begin{equation}
  \begin{split}
    V_\nu^{\rm LO}(\bm p',\bm p)&=\frac{M^2}{\sqrt{(\bm p'{}^2+M^2)(\bm p^2+M^2)}}V_\nu(\bm p',\bm p)\\
    &=V_\nu(\bm p',\bm p)\left(1-\frac{\bm p'{}^2+\bm p^2}{2M^2}+O\left(\frac{p^4}{M^4}\right)\right).
  \end{split}
\end{equation}
Here, $V_\nu(\bm p',\bm p)$ is the LO neutrino potential [Eq.~(\ref{eq.Vnu})].
After expanding the operator in terms of $1/M$, one finds the leading relativistic correction has the opposite sign from the nonrelativistic part.
The relativistic correction grows with increasing momentum cutoff and is significant at large cutoffs $\Lambda\gtrsim M$.

\section{Prediction of the $0\nu\beta\beta$ amplitude}
\begin{figure}[!htpb]
    \centering
	\includegraphics[width=0.45\textwidth]{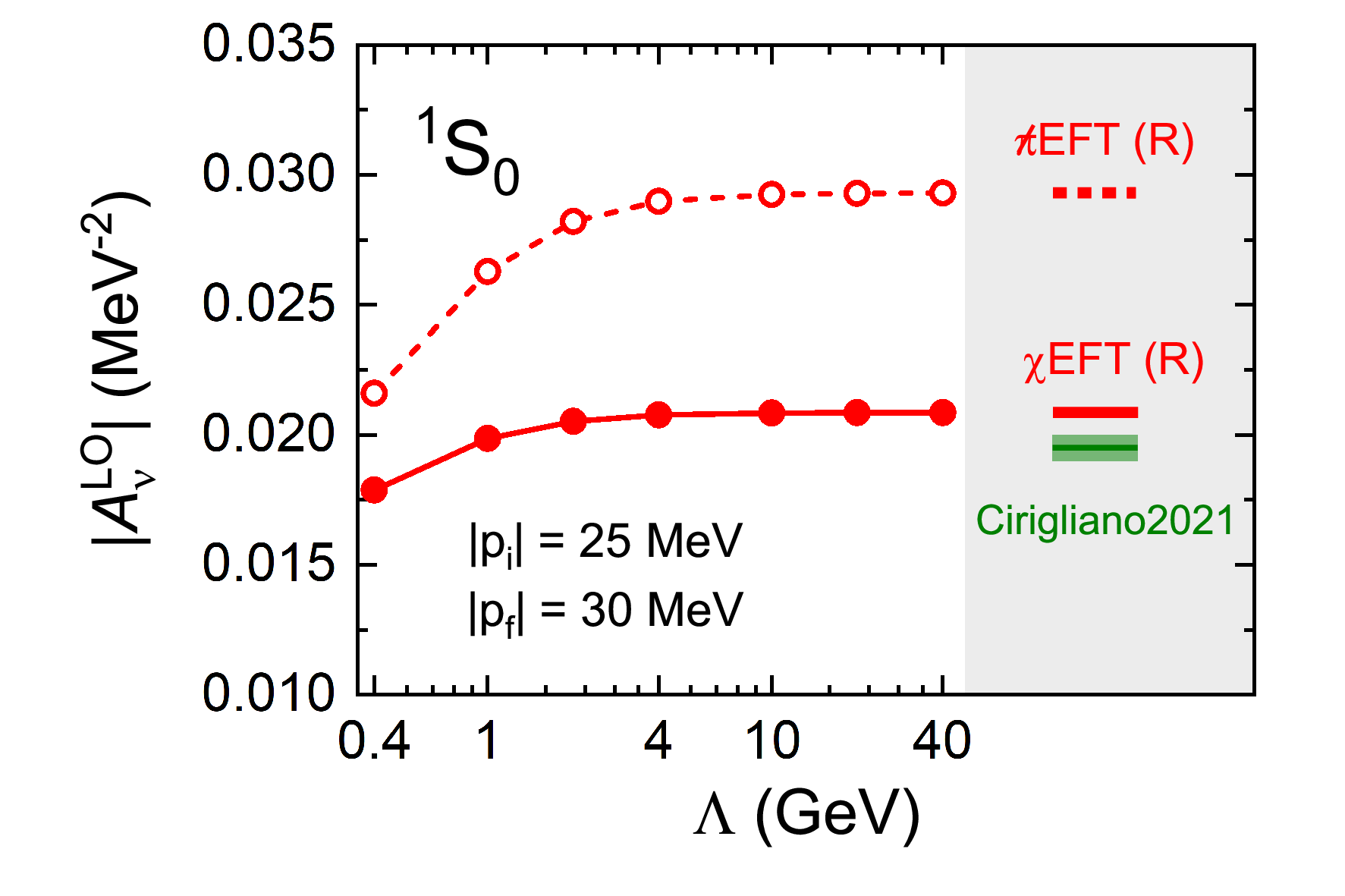}
	\caption{(Color online). The amplitude $\mathcal{A}^{\rm LO}_\nu$ for $|\bm p_i|=25$ MeV and $|\bm p_f|=30$ MeV given by the relativistic chiral and pionless EFTs, in comparison with the result estimated previously in Ref.~\cite{Cirigliano2021Phys.Rev.Lett.172002}.
	}\label{fig3}
\end{figure}
In Fig.~\ref{fig3}, we benchmark our calculated amplitude against the one estimated previously at the kinematics $|\bm p_i|=25$ MeV and $|\bm p_f|=30$ MeV~\cite{Cirigliano2021Phys.Rev.Lett.172002,Cirigliano2021J.HighEnergyPhys.289}.
Such a comparison is not trivial at all, because the two amplitudes are obtained in distinct approaches.
The amplitude obtained in this work is properly renormalized by the relativity and there is no need to introduce any unknown counter terms, while in the previous estimation~\cite{Cirigliano2021Phys.Rev.Lett.172002,Cirigliano2021J.HighEnergyPhys.289}, one has to renormalize the amplitude by introducing an unknown contact term, whose size was constrained by additional model-dependent inputs within a certain range.

It is remarkable that the renormalized amplitude obtained with the relativistic chiral EFT, $\mathcal{A}^{\rm LO}_\nu = -0.0209~\rm{MeV}^{-2}$, is quite consistent with the previous estimation~\cite{Cirigliano2021Phys.Rev.Lett.172002,Cirigliano2021J.HighEnergyPhys.289}, and their difference is only about $10\%$.
This demonstrates the validity of both approaches for $0\nu\beta\beta$ transitions, but the present relativistic framework avoids the uncertainties introduced by the model-dependent inputs.
In addition, the theoretical uncertainties in the present framework can be systematically improved by moving to higher orders of the chiral expansion.
The theoretical uncertainty originating from the truncation of the chiral expansion was rarely discussed in previous works on the $nn\rightarrow ppe^-e^-$ amplitude. Following Ref.~\cite{Epelbaum2015Eur.Phys.J.A53}, a very rough estimation could be given by $m_\pi/\Lambda_b$ with $\Lambda_b=600$ MeV the breakdown scale of chiral expansion, i.e., around $23\%$.

Compared to the relativistic chiral EFT, its pionless counterpart predicts an amplitude larger by about $40\%$.
We would expect that the predictions of the amplitudes with chiral and pionless EFTs should come closer at higher orders, as is the case for $NN$ scattering~\cite{Kaplan1996Nucl.Phys.B629659,Long2012Phys.Rev.C024001}, and it would be interesting to examine this in the future.

\begin{figure}[!htbp]
    \centering
	\includegraphics[width=0.45\textwidth]{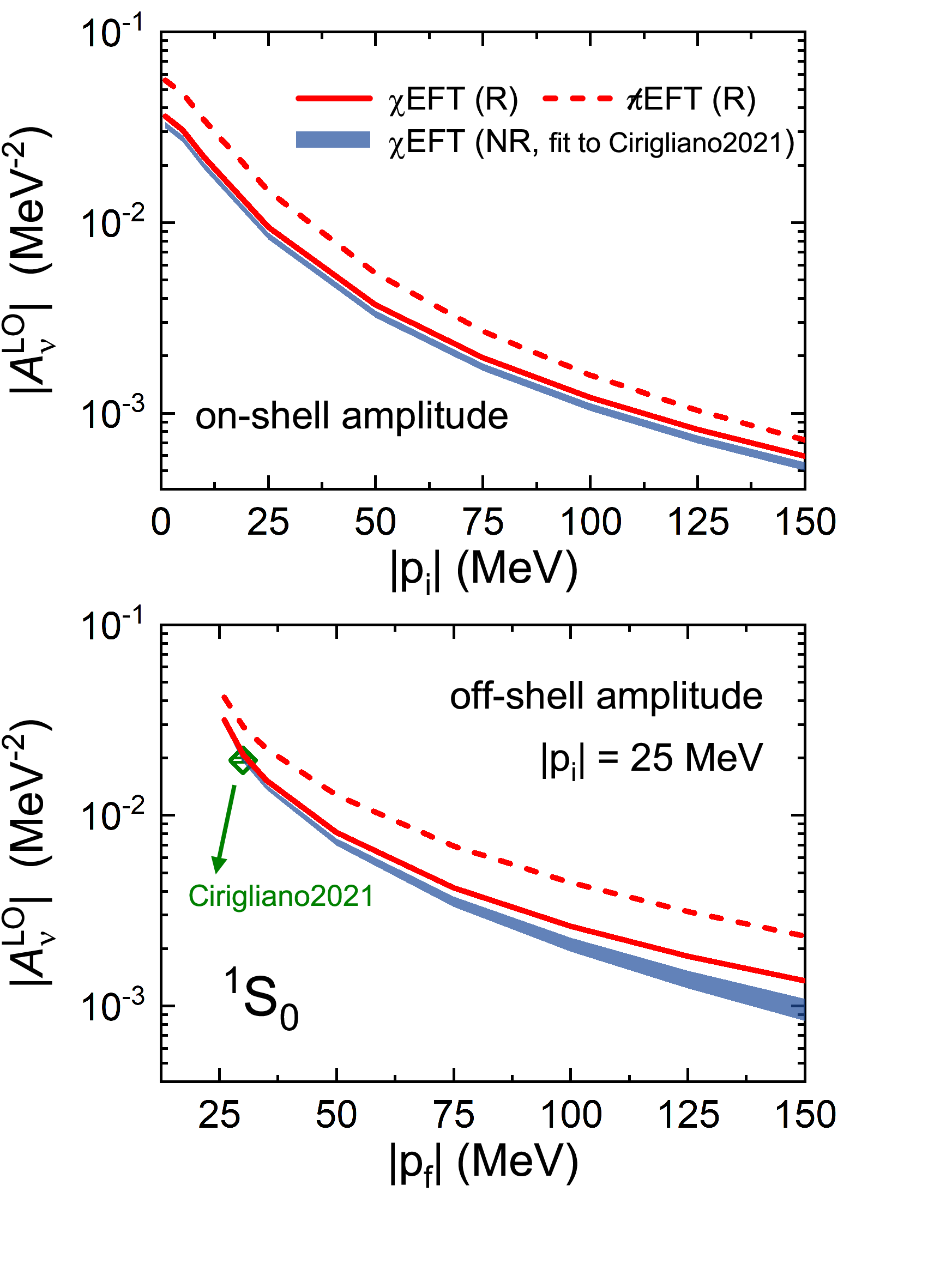}
	\caption{(Color online). The renormalized on-shell (upper) and off-shell (lower) amplitudes $\mathcal{A}^{\rm LO}_\nu$ at various kinematic points.
		The blue band represents the nonrelativistic results, in which the contact term is fitted to the synthetic datum (shown as diamond) for the amplitude at kinematics $|\bm p_i|=25$ MeV and $|\bm p_f|=30$ MeV, i.e., $\mathcal{A}^{\rm LO}_\nu = -0.0195(5)~\rm{MeV}^{-2}$~\cite{Cirigliano2021Phys.Rev.Lett.172002}. The width of the band reflects the uncertainty propagated from the synthetic datum.
}\label{fig4}
\end{figure}

Figure~\ref{fig4} depicts the renormalized on-shell and off-shell amplitudes at various kinematic points.
The results given by the relativistic pionless and chiral EFTs are compared with those given by the nonrelativistic chiral EFT, in which the contact term is fitted to the synthetic datum for the amplitude at kinematics $|\bm p_i|=25$ MeV and $|\bm p_f|=30$ MeV, i.e., $\mathcal{A}^{\rm LO}_\nu = -0.0195(5)~\rm{MeV}^{-2}$~\cite{Cirigliano2021Phys.Rev.Lett.172002}.
With the fitted contact term, the nonrelativistic amplitudes at other kinematic points can also be renormalized.
The uncertainty of the synthetic datum reflects its systematic error from the model dependences, which include the neglect of inelastic intermediate states, the parametrization of momentum dependence of nucleon-nucleon scattering amplitude, and the phenomenological form factors~\cite{Cirigliano2021Phys.Rev.Lett.172002,Cirigliano2021J.HighEnergyPhys.289}.
These uncertainties are in turn propagated to the contact term required in the nonrelativistic chiral EFT by fitting to the synthetic datum, and also the resultant amplitudes at all other kinematic points.
In contrast, the present relativistic framework provides a model-free prediction of the amplitudes, and they are generally consistent with the nonrelativistic estimations.
The relativistic results are slightly larger than the nonrelativistic ones by about 10$\%$-20$\%$ for the on-shell amplitude, while for the off-shell amplitude, the enhancement becomes more sizable especially at high final momentum, e.g., about $40\%$ at $|\boldsymbol p_f|=150$ MeV.
This should be understandable because the contact term in the nonrelativistic framework is determined at extremely low-energy kinematics.

In the present relativistic framework, the contact term is expected to appear at next-to-next-to-leading order, so its contribution is significantly suppressed as compared to the nonrelativistic case.
Therefore, the present results are model-free at LO, in the sense that the present approach is free of model-dependent inputs beyond the framework of chiral EFT.
There exist some variants of power counting schemes in both nonrelativistic and relativistic EFTs~\cite{Hammer2020Rev.Mod.Phys.025004}, but it does not preclude the chiral EFT from providing model-independent results for physical observables.
Such scheme dependence could in principle be controlled by the EFT paradigm within the truncation errors.

The present work provides the opportunity for a direct benchmark between the chiral EFT predictions and the lattice QCD calculations for $0\nu\beta\beta$ decay, which is currently only possible at unphysically heavy quark masses due to the computational cost~\cite{Shanahan2017Phys.Rev.Lett.062003,Tiburzi2017Phys.Rev.D054505,Davoudi2021Phys.Rep.174}.
Extending the present framework to meet the lattice QCD conditions is straightforward, since it only takes the axial coupling constant $g_A$ and the $^1S_0$ scattering length $a_{np}$ as inputs.
A direct comparison against the upcoming lattice QCD simulations for the $nn\rightarrow ppe^-e^-$ amplitudes, instead of using them as inputs, will provide a stringent test on the validity of the present framework.

The present results can help to address the uncertainty of nuclear matrix elements in nuclear-structure calculations for isotopes of experimental interest for $0\nu\beta\beta$ searches.
On the one hand, the present amplitudes can serve as alternative synthetic data to determine the contact term in nonrelativistic nuclear-structure calculations.
Note that the contact-term effects could be amplified in realistic $0\nu\beta\beta$ transitions~\cite{Cirigliano2018Phys.Rev.Lett.202001}.
On the other hand, a consistent relativistic {\it ab initio} calculation of nuclear $0\nu\beta\beta$ decay rates starting from relativistic chiral forces is highly demanded in view of the recent significant progress in both relativistic {\it ab initio} methods~\cite{Shen2016Chin.Phys.Lett.102103,Wang2021Phys.Rev.C054319,Yang2022Phys.Lett.B137587} and nuclear forces
~\cite{Epelbaum2015Eur.Phys.J.A71,Ren2022Phys.Rev.C034001,Ren2018Chin.Phys.C014103,Lu2022Phys.Rev.Lett.142002}.

It is interesting to extend the present framework to investigate other processes related to $0\nu\beta\beta$ decay, e.g., the two-neutrino double-beta decay and the pion double charge exchange reactions.
For the cases of these two processes, the strong interactions in the intermediate state should be taken into account at LO, in contrast to the case of $0\nu\beta\beta$ decay.
As a result, the renormalization-group analysis for the two-nucleon amplitudes for these processes could be more involved,
due to the tensor forces generated by the one-pion exchange in the intermediate states.

\section{Summary and Outlooks}
We present a model-free prediction of the $nn\rightarrow pp e^-e^-$ amplitude by developing a relativistic framework based on chiral EFT.
Contrary to the nonrelativistic case, we show that the amplitude can be renormalized at LO without any uncertain contact operators.
The calculated amplitude is slightly larger than the previous model-dependent estimation at the kinematics $|\bm p_i|=25$ MeV and $|\bm p_f|=30$ MeV~\cite{Cirigliano2021Phys.Rev.Lett.172002,Cirigliano2021J.HighEnergyPhys.289} by about 10\%, providing a highly nontrivial validation for the previous result.
The enhancement of the amplitude could be more sizable for either off-shell amplitude or high-momentum kinematics.
Therefore, the present amplitude defines a stringent benchmark for the previously estimated amplitude, and can help to address the uncertainty of the nuclear matrix elements in nuclear-structure calculations for isotopes of experimental interest for $0\nu\beta\beta$ searches.
In addition, the present work motivates a relativistic {\it ab initio} calculation of $0\nu\beta\beta$ decay in light and medium-mass nuclei.

\section*{Acknowledgments}
We thank Xu Feng, Lisheng Geng, Teng Wang, and Yakun Wang for stimulating discussions at various stages of this work.
This work has been supported in part by the National Natural Science Foundation of China (Grants No. 12141501, No. 123B200320, No. 12070131001, No. 11935003, No. 11975031), and the High-performance Computing Platform of Peking University.
We acknowledge the funding support from the State Key Laboratory of Nuclear Physics and Technology, Peking University (Grant No. NPT2023ZX03).


\end{document}